\documentclass[aps,floatfix,showpacs]{revtex4}
\usepackage{graphics,epsfig}
\begin{document}
\title{Differential Invariants of the Kerr Vacuum}
\author{Kayll Lake \cite{email}}
\affiliation{Department of Physics and Department of Mathematics
and Statistics, Queen's University, Kingston, Ontario, Canada, K7L
3N6 }
\date{\today}
\begin{abstract}
The norms associated with the gradients of the two
non-differential invariants of the Kerr vacuum are examined.
Whereas both locally single out the horizons, their global
behavior is more interesting. Both reflect the background angular
momentum as the volume of space allowing a timelike gradient
decreases with increasing angular momentum becoming zero in the
degenerate and naked cases. These results extend directly to the
Kerr-Newman geometry.
\end{abstract}
\maketitle

Recently \cite{lake} I reviewed, in the Kerr vacuum, the two
independent invariants derivable from the Weyl tensor (and its
dual) without differentiation, and showed that both of these
non-differential invariants must be examined in order to avoid an
erroneous conclusion that the ring singularity of this spacetime
is, in any sense, ``directional". In this companion piece I
examine differential invariants with one differentiation (that is,
invariants containing no more than three derivatives of the metric
tensor). After reviewing invariants constructed directly from
covariant derivatives of the Weyl tensor (and its dual),
invariants which appear to have little if any significance, I
explore the norms associated with the gradients of the two
non-differential invariants. These invariants do single out the
horizons, but their global behavior is more interesting as they
reflect the background angular momentum in the following way: the
volume of space allowing a timelike gradient decreases with
increasing angular momentum becoming zero in the degenerate case.
The nakedly singular cases allow no timelike gradients. Important
multimedia enhancements to this work are at
\texttt{http://grtensor.org/diffweyl/}

\bigskip
We start by reviewing invariants constructed directly from
covariant derivatives of the Weyl tensor. In terms of the familiar
Boyer-Lindquist coordinates \cite{boyer}, write $\textit{x} \equiv
r/a$ and $A \equiv m/a >0$, both with $a \neq 0$, and write $C_{i
j k l}$ as the Weyl tensor with $\bar{C}_{i j k l}$ its dual. It
follows that $\nabla_{m}C_{i j k l} \nabla^{m}{C}^{i j k
l}=-\nabla_{m} \bar{C}_{i j k l} \nabla^{m}\bar{C}^{i j k l}$ and
so there are two quadratic invariants derivable from covariant
derivatives of the Weyl tensor. Write these as $\mathcal{D}_s
\equiv \nabla_{m}C_{i j k l} \nabla^{m} \bar{C}^{i j k
l}m^6A^8/5760$ and $\mathcal{D} \equiv \nabla_{m}C_{i j k l}
\nabla^{m} C^{i j k l }m^6A^8/720$. It follows that
\begin{equation}
\mathcal{D}_s=-{\frac {x{\it x1}\, ( x-{\it x1} ) ( x+{\it x1}
 )  ( x-{\it x2} )  ( x+{\it x2} )
 ( x-{\it x3} )  ( x+{\it x3} )  f(x) }{ ( {x}^{2}+{{\it x1}}^{2 } ) ^{9}}},
\end{equation}
and
\begin{equation}
\mathcal{D}={\frac { ( x-{\it x6} )  ( x+{\it x6} ) ( x-{\it x7} )
( x+{\it x7} ) ( x-{\it x8}
 )  ( x+{\it x8} )  ( x-{\it x9} )
 ( x+{\it x9} )  f(x) }{ ( {x}^{2}+{{\it x1}}^{2} ) ^{9}}},
\end{equation}
where
\begin{equation}
{\it x1}=\cos ( \theta ),\;\;\;{\it x2}=( 1-\sqrt {2} ){\it
x1},\;\;\;{\it x3}=( 1+\sqrt {2} ){\it x1},
\end{equation}
and
\begin{equation}
f(x)\equiv{x}^{2}-2\,{\frac {x}{A}}+{\it x1} ^{2} ,
\end{equation}
which allows the restricted factorization
\begin{equation}
 ( x-{\it x4} )  ( x-{\it
x5} ),
\end{equation}
with
\begin{equation}
{\it x4}=\,{\frac {1+\,\sqrt {1-(A{\it x1}) ^{2}}}{A}},\;\;\;{\it
x5}=\,{\frac {1-\,\sqrt {1-(A{\it x1})^{2}}}{A}},
\end{equation}
for $(A{\it x1}) ^{2} \leq 1$,
\begin{equation}
{\it x6}= ( -1-\sqrt {2}+\sqrt {4+2\,\sqrt {2}} ) {\it
x1},\;\;\;{\it x7}= ( 1-\sqrt {2}+\sqrt {4-2\,\sqrt {2}} ) {\it
x1},
\end{equation}
and
\begin{equation}
{\it x8}=   ( -1+\sqrt {2}+\sqrt {4-2\,\sqrt {2}} ) {\it
x1},\;\;\; {\it x9}= ( 1+\sqrt {2}+\sqrt {4+2\,\sqrt {2}} ) {\it
x1}.
\end{equation}

\bigskip
The invariant $\mathcal{D}$ was first discussed by Karlhede et al
\cite{kla} over two decades ago, but an unfortunate factorization
in that paper would lead to the suggestion that $\mathcal{D}$
vanishes only on the ergo surfaces ${\it x4}$ and ${\it x5}$. (An
extensive discussion of $\textit{x4}$ is given in \cite{pelavas}.)
The roots for $\mathcal{D}$ have been given previously by Gass et
al \cite{gass}. The significance of $\mathcal{D}_s$, if any, is
unclear. The significance of $\mathcal{D}$ also remains unclear
except for the fact that in the equatorial plane (${\it x1}=0,
{\it x}>0$) it vanishes only on the outer ergo surface (${\it
x4}$, ${\it x}=2/A$). What is clear is that in the limit $x
\rightarrow 0,\; \theta \rightarrow \pi/2$ ($\equiv \mathcal{S}$)
both $\mathcal{D}_s$ and $\mathcal{D}$ remain zero along the inner
ergo surface (${\it x}={\it x5} \rightarrow 0$). Three dimensional
images of $\mathcal{D}_s$ and $\mathcal{D}$ are available at the
web site.

\bigskip
The Kerr vacuum has two independent invariants derivable from the
Riemann tensor without differentiation \cite{lake}. The norms
associated with the gradients of these scalar fields form a rather
natural class of differential invariant for investigation and yet
there appears to be no discussion of these in the literature.
These gradients are defined by $v_{e} \equiv \nabla_{e} (C_{i j k
l} C^{i j k l})=-\nabla_{e} (\bar{C}_{i j k l} \bar{C}^{i j k l})$
and $w_{e} \equiv \nabla_{e} (C_{i j k l} \bar{C}^{i j k
l})=\nabla_{e} (\bar{C}_{i j k l} C^{i j k l})$. For $v$ a direct
calculation gives
\begin{equation}
v_{r}=-288\,{\frac {x ( -7\, ( {\it x1}
 ) ^{2} ( -5\, ({\it x1}) ^
{2}{x}^{2}+ ( {\it x1}) ^{4}+3\,{x}^{4}
 ) +{x}^{6} ) }{{m}^{5}{A}^{7} ( {x}^{2}+ ( {\it x1}) ^{2} )
 ^{7}}},\;\;v^{r}={\frac {{x}^{2}A-2\,x+A}{A ( {x}^{2}+{{\it x1}}^{2}) }}v_{r}
,\end{equation}
\begin{equation}
v_{\theta}=-288\,{\frac {{\it x1}  ( - ( {\it x1}) ^{2} ( -21\, (
{\it x1}  ) ^{2}{x}^{2}+ ({\it x1}
 ) ^{4}+35\,{x}^{4} ) +7\,{x}^{6} ) }{{m}^{5}{A}^{7}
 ( {x}^{2}+ ( {\it x1}) ^{2}
 ) ^{7}}}
,\;\; v^{\theta}={\frac {1-{{\it x1}}^{2}}{{x}^{2}+{{\it
x1}}^{2}}}v_{\theta},
\end{equation}
with $v_{\phi}=v^{\phi}=v_{t}=v^{t}=0$. Similarly for $w$,
$w_{r}=-v_{\theta}$, $w_{\theta}=v_{r}$ ,
$w^{r}=(x^2A-2x+A)/(A(x^2+{\it x1}^2))w_{r}$, $w^{\theta}=(1-{\it
x1}^2)/(x^2+x1^2)w_{\theta}$ and
$w_{\phi}=w^{\phi}=w_{t}=w^{t}=0$.

\bigskip
The scaled norm $\mathcal{V} \equiv
(v_{e}v^{e}){A}^{15}{m}^{10}/288^2$ is given by
\begin{eqnarray}
\mathcal{V}= -(-{x}^{14}( {x}^{2}A-2\,x+A) +{{\it x1}}^{2} (
A{{\it x1}}^{14}-42\,{x}^{2}A{{\it x1}}^{12}-{{\it x1}}^{12}A+
462\,{x}^{4}A{{\it x1}}^{10}+98\,{x}^{3}{{\it
x1}}^{10}-7\,{x}^{2}{{ \it
x1}}^{10}A\\\nonumber-994\,{x}^{6}A{{\it
x1}}^{8}-980\,{x}^{5}{{\it x1}}^{8}- 21\,{x}^{4}{{\it
x1}}^{8}A+3038\,{x}^{7}{{\it x1}}^{6}-35\,{x}^{6}{{ \it
x1}}^{6}A+994\,{x}^{10}A{{\it x1}}^{4}-2968\,{x}^{9}{{\it x1}}^{4}
-35\,{x}^{8}{{\it x1}}^{4}A\\\nonumber-462\,{x}^{12}A{{\it
x1}}^{2}+1022\,{x}^{11 }{{\it x1}}^{2}-21\,{x}^{10}{{\it
x1}}^{2}A+42\,{x}^{14}A-84\,{x}^{13} -7\,{x}^{12}A))\\\nonumber/(
{x}^{2}+{{\it x1}}^{2}) ^{15} .
\end{eqnarray}
Along ${\it x}=0$
\begin{equation}
\mathcal{V}=-{\frac {A( {\it x1}-1) ( {\it x1}+1) }{{{ \it
x1}}^{16}}}.
\end{equation}
In the equatorial plane ${\it x1}=0$
\begin{equation}
\mathcal{V}={\frac {{x}^{2}A-2\,x+A}{{x}^{16}}},
\end{equation}
and along the axis ${\it x1}^2=1$
\begin{equation}
\mathcal{V}={\frac {{x}^{2}( {x}^{6}-21\,{x}^{4}+35\,{x}^{2}-7)
^{2} ( {x}^{2}A-2\,x+A ) }{( {x}^{2}+1 ) ^{15}}}.
\end{equation}
For $0<A \leq 1$, ${x}^{2}A-2\,x+A=0$ at the horizons
\begin{equation}
{\it x10}={\frac {1+\sqrt {1-{A}^{2}}}{A}},\;\;\;{\it x11}={\frac
{1-\sqrt {1-{A}^{2}}}{A}}.
\end{equation}
The real positive roots to $x^6-21x^4+35x^2-7=0$ are ${\it x}
\simeq 0.4815746188, 1.253960338, 4.381286268$. Unlike
$\mathcal{D}$ or $\mathcal{D}_{s}$, $\mathcal{V}$ shows a strong
sign dependence with $A$. Whereas the vanishing of $\mathcal{V}$
in the equatorial plane (and on the axis away from the discrete
roots given) obviously singles out the horizons, the global
behavior of $\mathcal{V}$ is rather more interesting than that. In
general terms, the area of regions where $\mathcal{V}<0$ (and so
$v$ is timelike) on any $r - \theta$ hypersurface decreases as $A
\rightarrow 1^{-}$ and vanishes for $A \geq 1$. This is shown in
Figures 1 through 3 where plots of $\mathcal{V}$ are shown
truncated at $ \mathcal{V}=0,+1$ for $A=1/2, 0.95, 1$. An
animation of this evolution is at the web site. A plot of $
\mathcal{V}=0$ for $A$ close to 1 is shown in Figure 4.

\bigskip
The scaled norm $\mathcal{W} \equiv
(w_{e}w^{e}){A}^{15}{m}^{10}/288^2$ is given by
\begin{eqnarray}
\mathcal{W}=(A{x}^{14}+48\,A{{\it x1}}^{2}{x}^{14}-98\,{{\it
x1}}^{2}{x}^{ 13}-448\,A{{\it x1}}^{4}{x}^{12}+7\,A{{\it
x1}}^{2}{x}^{12}+980\,{{ \it x1}}^{4}{x}^{11}+21\,A{{\it
x1}}^{4}{x}^{10}+1008\,A{{\it x1}}^{6}
{x}^{10}\\\nonumber-3038\,{{\it x1}}^{6}{x}^{9}+35\,A{{\it
x1}}^{6}{x}^{8}+2968\, {{\it x1}}^{8}{x}^{7}-1008\,A{{\it
x1}}^{10}{x}^{6}+35\,A{{\it x1}}^{8 }{x}^{6}-1022\,{{\it
x1}}^{10}{x}^{5}+21\,A{{\it x1}}^{10}{x}^{4}\\\nonumber+448
\,A{{\it x1}}^{12}{x}^{4}+84\,{{\it x1}}^{12}{x}^{3}+7\,A{{\it
x1}}^{ 12}{x}^{2}-48\,A{{\it x1}}^{14}{x}^{2}-2\,{{\it
x1}}^{14}x+A{{\it x1}} ^{14})\\\nonumber/( {x}^{2}+{{\it x1}}^{2})
^{15}.
\end{eqnarray}
Along ${\it x}=0$
\begin{equation}
\mathcal{W}={\frac {A}{{{\it x1}}^{16}}}.
\end{equation}
In the equatorial plane ${\it x1}=0$
\begin{equation}
\mathcal{W}={\frac {A}{{x}^{16}}},
\end{equation}
and along the axis ${\it x1}^2=1$
\begin{equation}
\mathcal{W}={\frac {( 7\,{x}^{6}-35\,{x}^{4}+21\,{x}^{2}-1) ^{2} (
A{x}^{2}-2\,x+A) }{( {x}^{2}+1) ^{15}}}.
\end{equation}
The real positive roots to $7\,{x}^{6}-35\,{x}^{4}+21\,{x}^{2}-1
=0$ are ${\it x} \simeq 0.2282434744, 0.7974733889, 2.076521397$.
Again, unlike $\mathcal{D}$ or $\mathcal{D}_{s}$, $\mathcal{W}$
shows a strong sign dependence with $A$. Whereas the vanishing of
$\mathcal{W}$ on the axis obviously singles out the horizons (away
from the discrete roots given), the global behavior of
$\mathcal{W}$, like $\mathcal{V}$, is rather more interesting than
that. In general terms, the area of regions where $\mathcal{W}<0$
(and so $w$ is timelike) on any $r - \theta$ hypersurface
decreases as $A \rightarrow 1^{-}$ and vanishes for $A \geq 1$ in
a manner similar to $\mathcal{V}$. This is shown in Figures 5
through 7 where plots of $\mathcal{W}$ are shown truncated at $
\mathcal{W}=0,+1$ for $A=1/2, 0.95, 1$. An animation of this
evolution is at the web site. A plot of $ \mathcal{W}=0$ for $A$
close to 1 is shown in Figure 8.

\bigskip
For completeness we note that the scaled norm $\mathcal{X} \equiv
(v_{e}w^{e}){A}^{16}{m}^{10}/288^2$ is given by
\begin{eqnarray}
\mathcal{X}=-{\frac {x{\it x1}\,( {x}^{6}-21\,{x}^{4}{{\it
x1}}^{2}+35\,{x}^ {2}{{\it x1}}^{4}-7\,{{\it x1}}^{6})(
7\,{x}^{6}-35\,{x} ^{4}{{\it x1}}^{2}+21\,{x}^{2}{{\it
x1}}^{4}-{{\it x1}}^{6})f(x) }{ ( {x}^{2}+{{\it x1}}^{2}) ^{15}}}.
\end{eqnarray}
$\mathcal{X}$ shares none of the interesting global properties of
$\mathcal{V}$ or $\mathcal{W}$.

\bigskip
The norms associated with the gradients of the two
non-differential invariants of the Kerr vacuum have been examined.
It has been shown that both locally single out the horizons, but
more importantly, their global behavior reflects the background
angular momentum as the volume of space allowing a timelike
gradient decreases with increasing angular momentum, becoming zero
in the degenerate and naked cases. One would certainly like to
know if these results are robust as regards an extension to the
Kerr-Newman geometry. In a subsequent note \cite{lake2} I show
that indeed they are.

\begin{acknowledgments}
This work was supported by a grant from the Natural Sciences and
Engineering Research Council of Canada. Portions of this work were
made possible by use of \textit{GRTensorII} \cite{grt}.
\end{acknowledgments}

\newpage
\begin{figure}
\epsfig{file=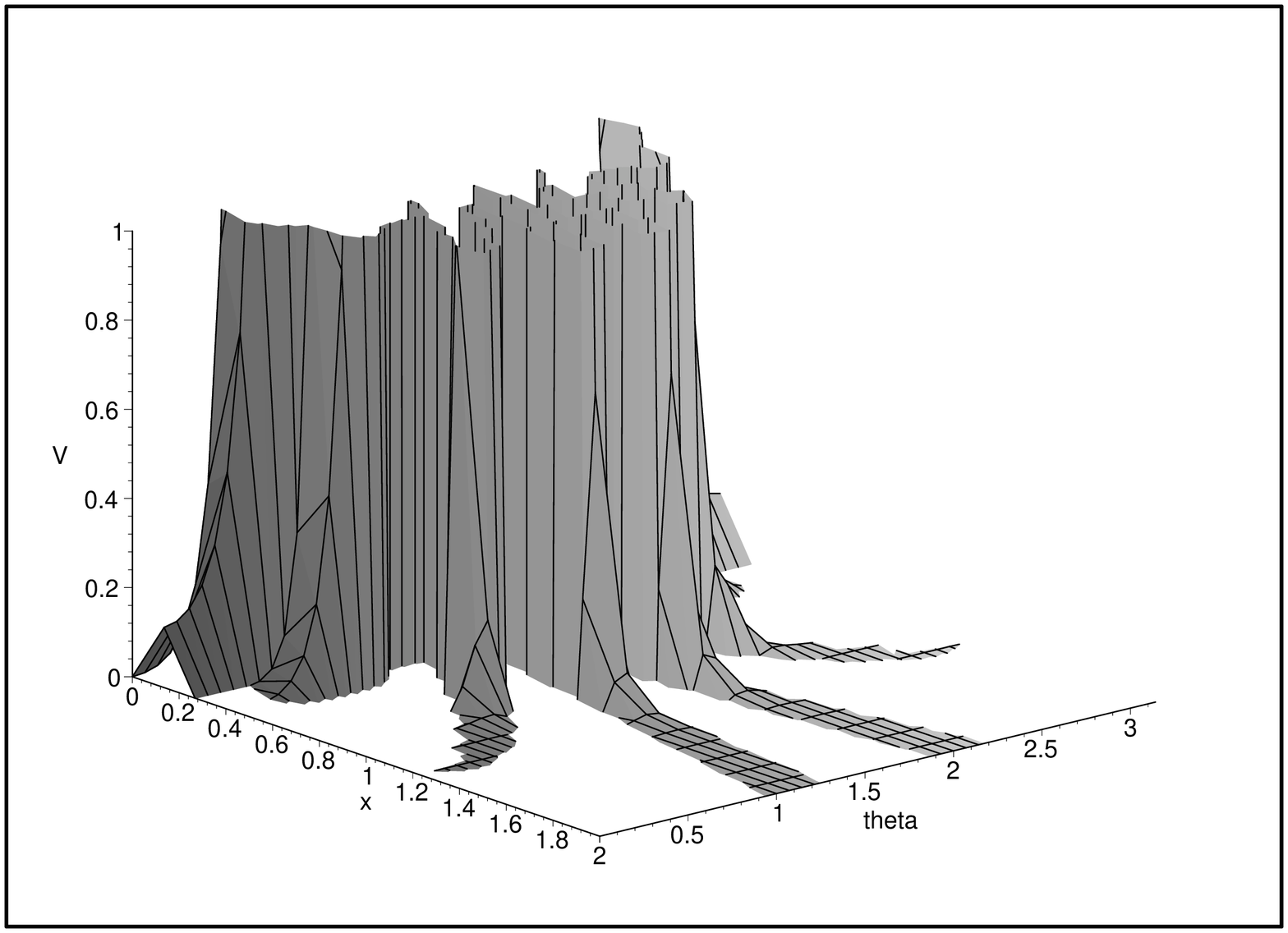,height=6in,angle=-90} \caption{Plot of
$ \mathcal{V}$ truncated at  $ \mathcal{V}=0,+1$ for $A=1/2$.}
\end{figure}

\newpage
\begin{figure}
\epsfig{file=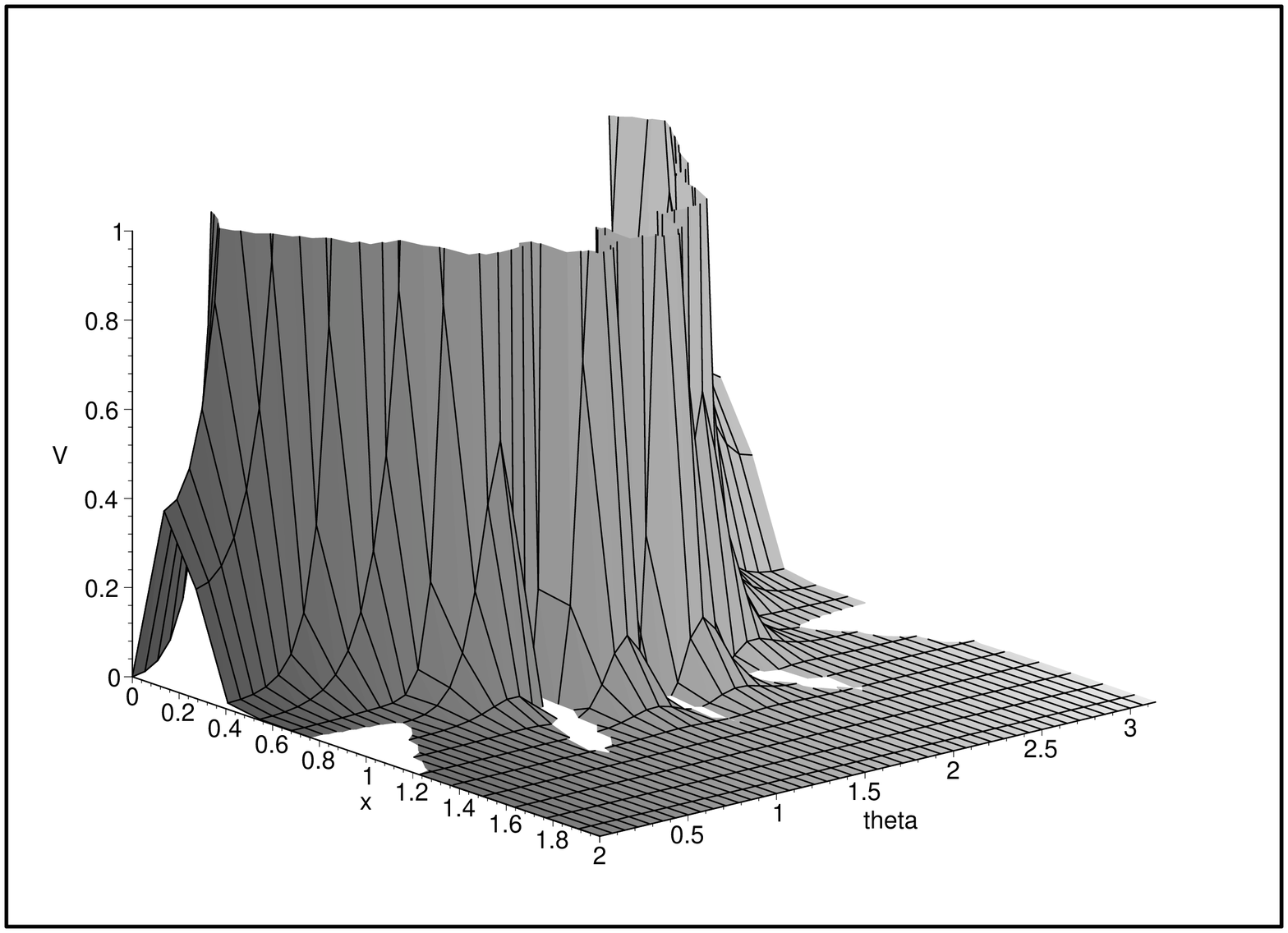,height=6in,angle=-90} \caption{Plot of $
\mathcal{V}$ truncated at  $ \mathcal{V}=0,+1$ for $A=0.95$.}
\end{figure}

\newpage
\begin{figure}
\epsfig{file=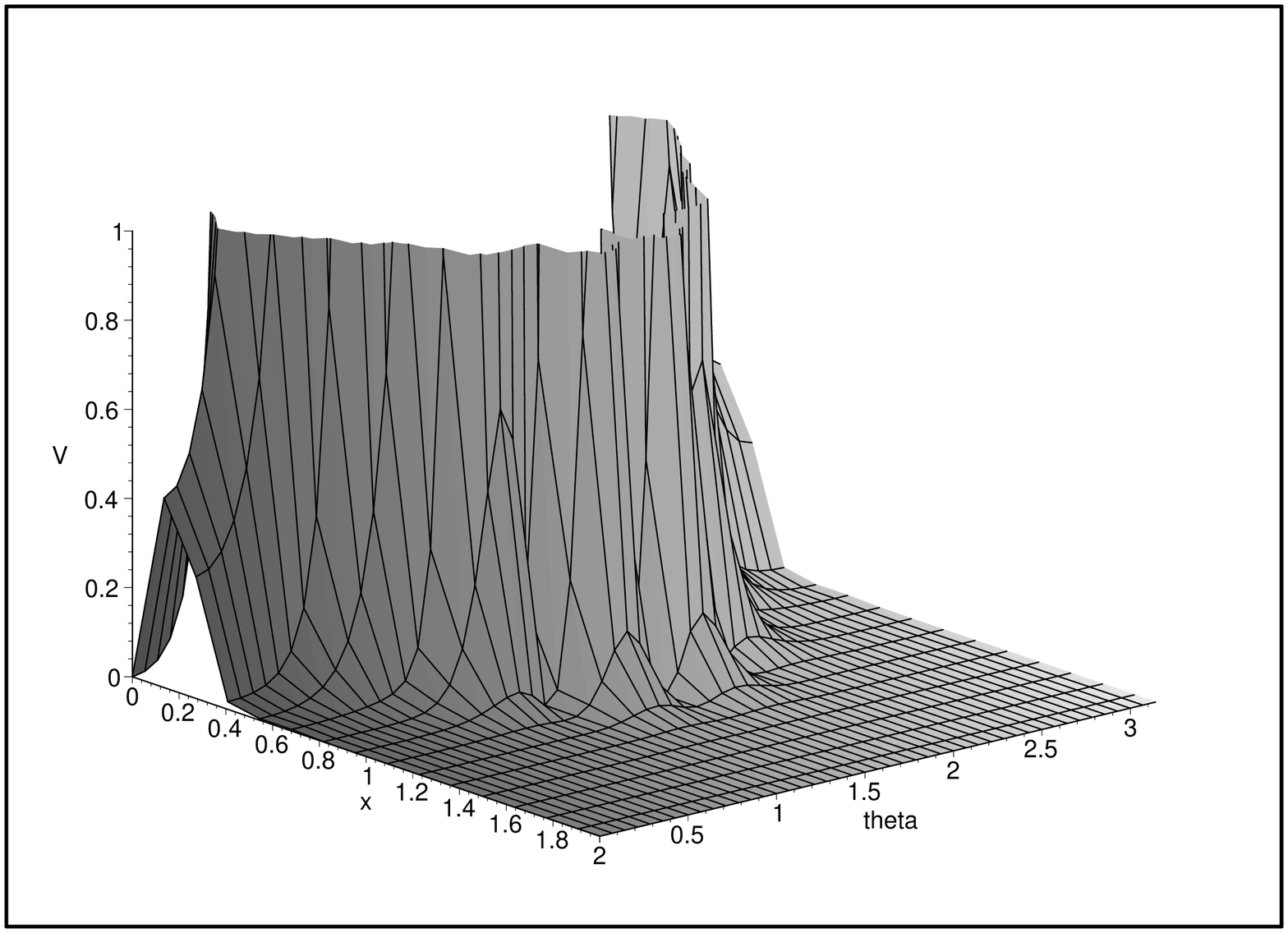,height=6in,angle=-90} \caption{Plot of $
\mathcal{V}$ truncated at  $ \mathcal{V}=0,+1$ for $A=1$.}
\end{figure}

\newpage
\begin{figure}
\epsfig{file=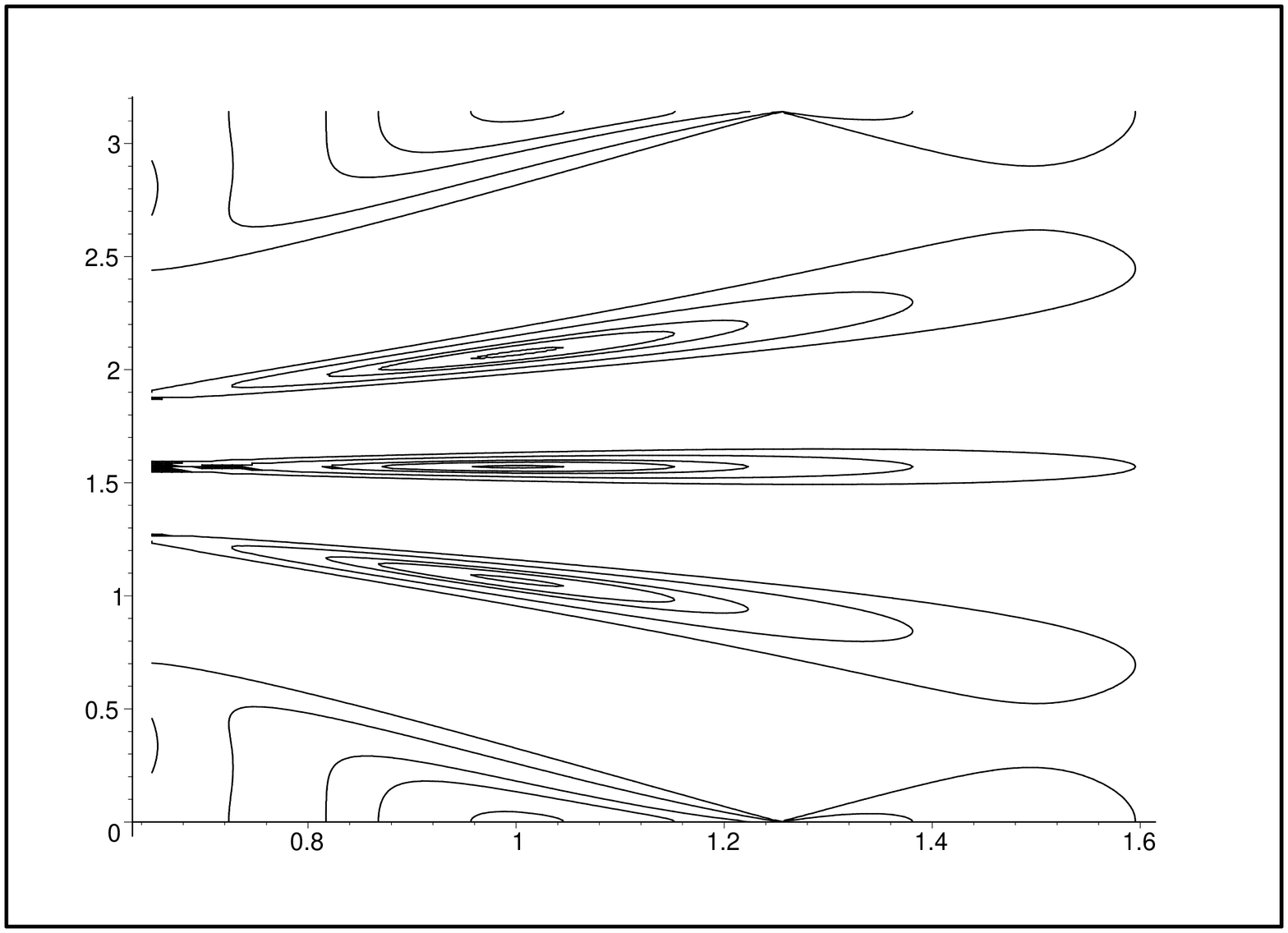,height=6in,angle=-90} \caption{Plot of
$ \mathcal{V}=0$ for $A=0.9, 0.95, 0.98, 0.99, 0.999$. Abscissa is
${\it x}$ and ordinate is $\theta$.  The enclosed area decreases
for increasing $A$ and vanishes for $A \geq 1$.}
\end{figure}

\newpage
\begin{figure}
\epsfig{file=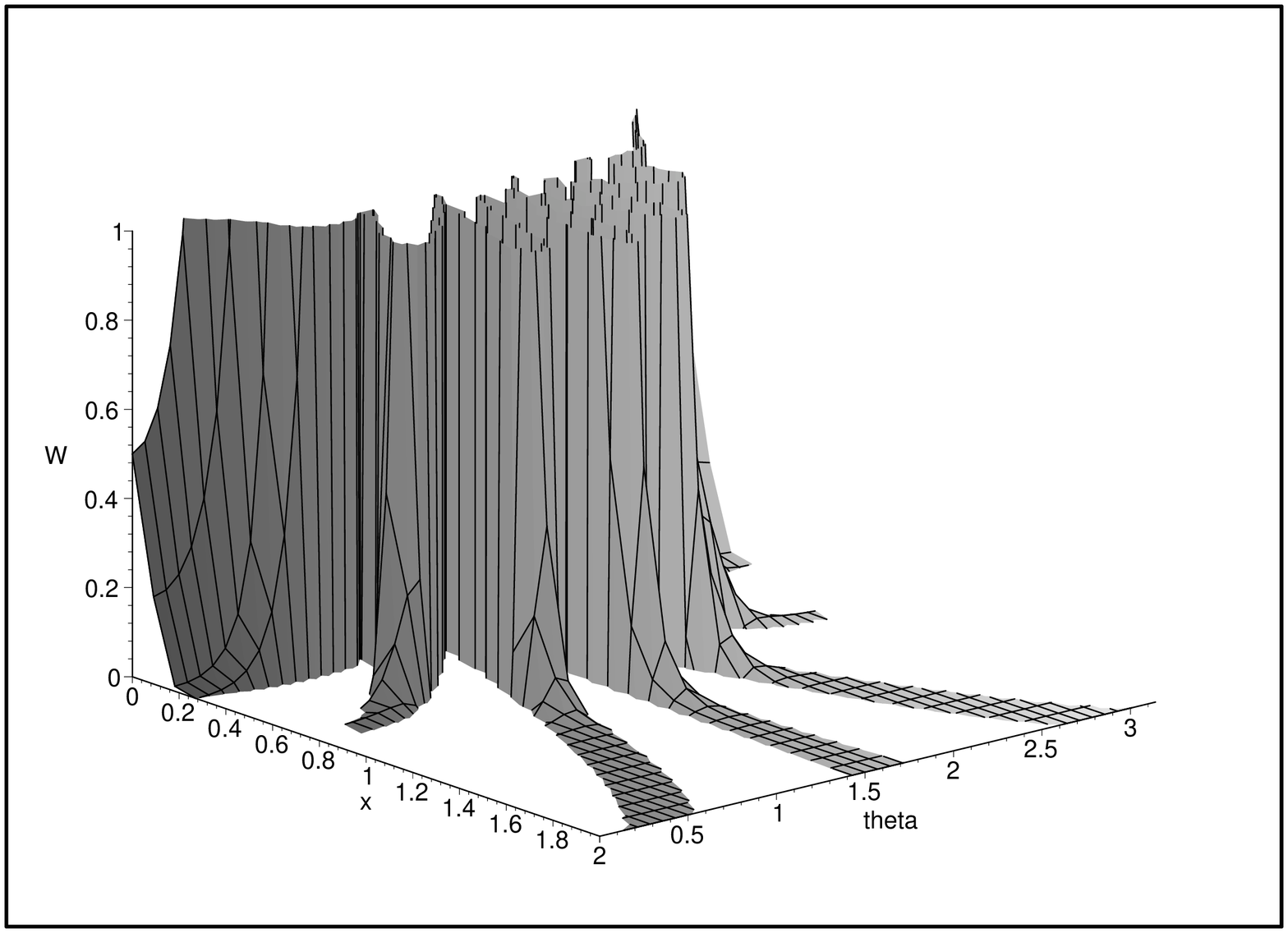,height=6in,angle=-90} \caption{Plot of
$ \mathcal{W}$ truncated at  $ \mathcal{W}=0,+1$ for $A=1/2$.}
\end{figure}

\newpage
\begin{figure}
\epsfig{file=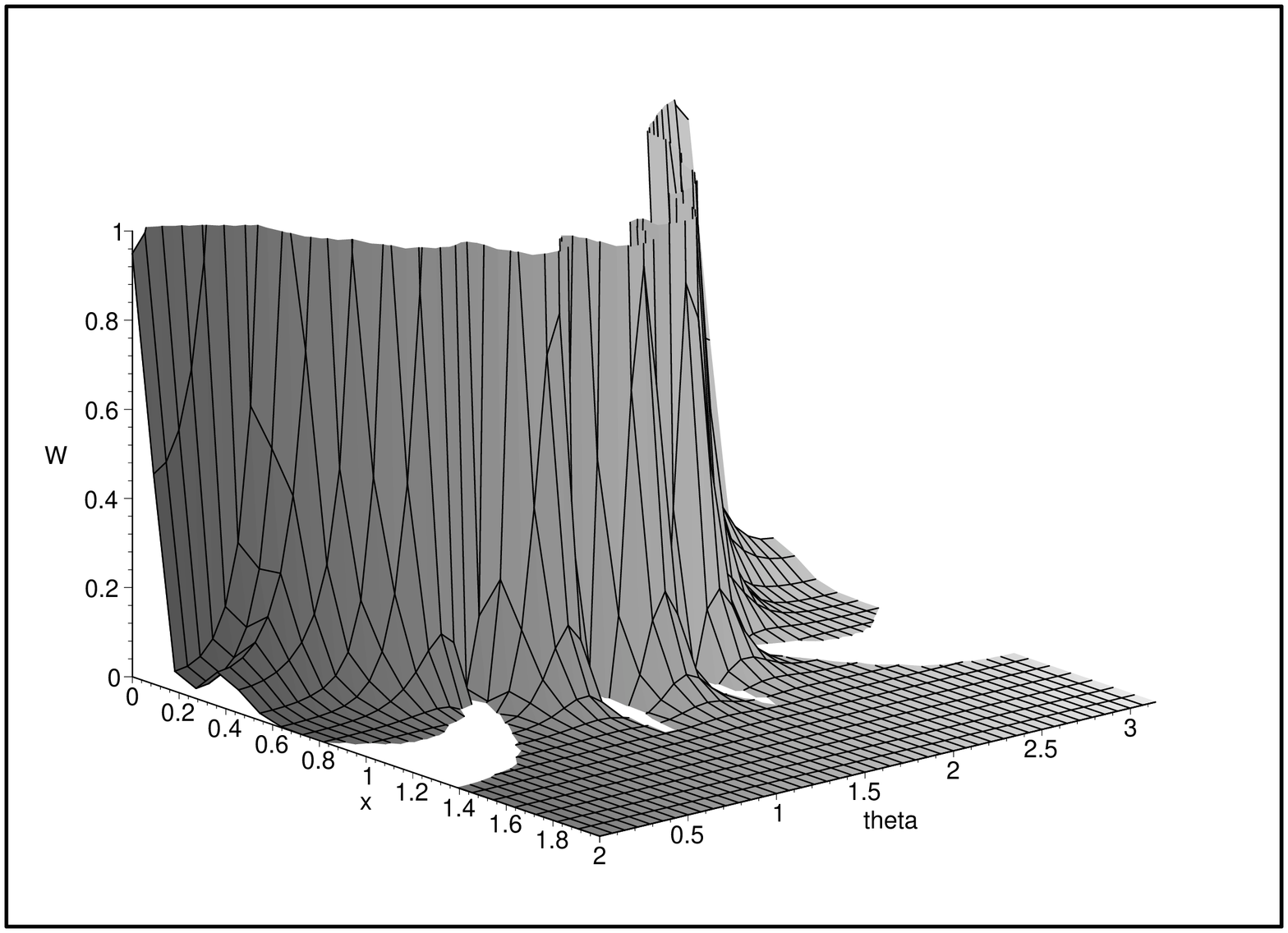,height=6in,angle=-90} \caption{Plot of $
\mathcal{W}$ truncated at  $ \mathcal{W}=0,+1$ for $A=0.95$.}
\end{figure}

\newpage
\begin{figure}
\epsfig{file=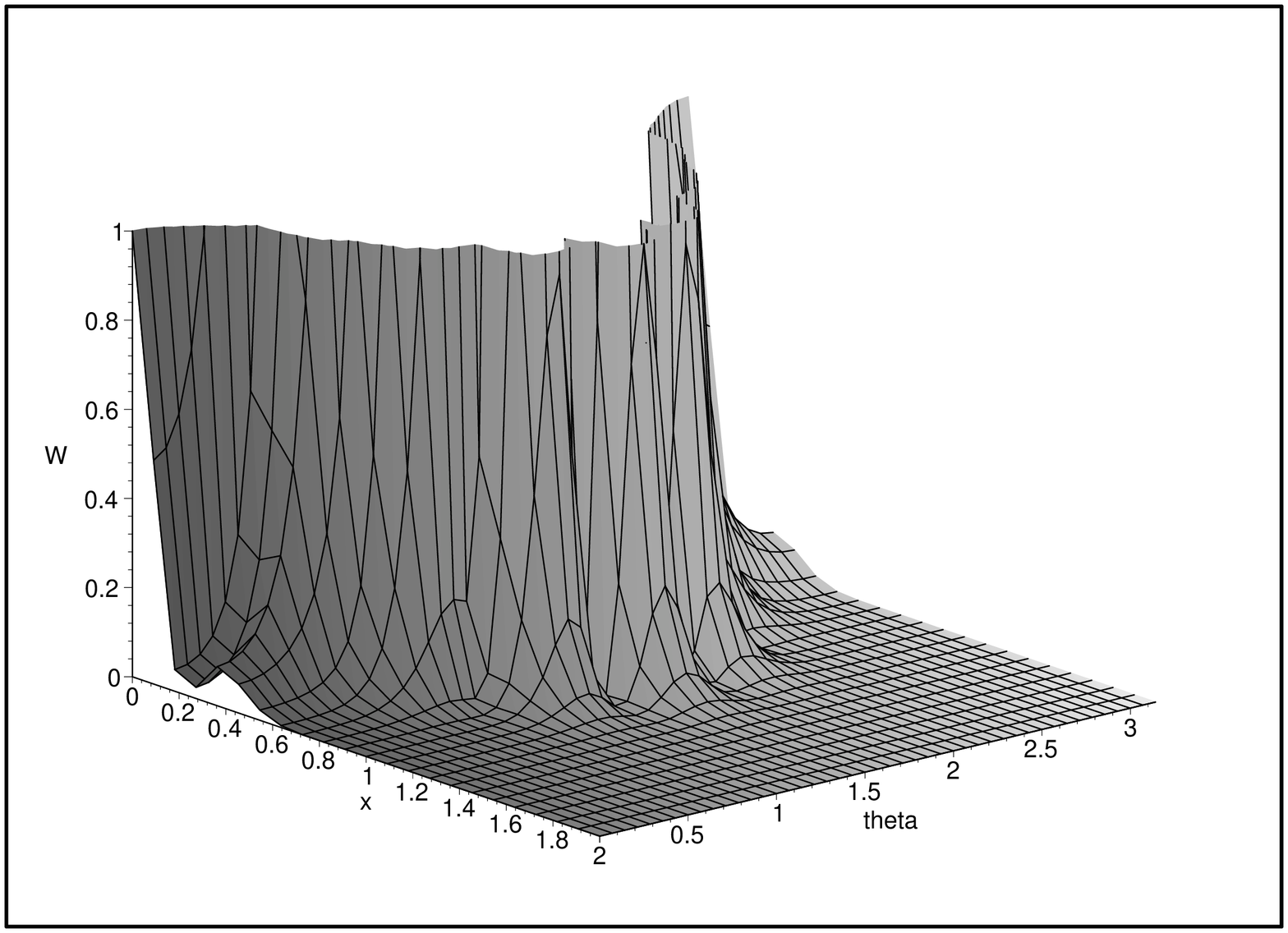,height=6in,angle=-90} \caption{Plot of $
\mathcal{W}$ truncated at  $ \mathcal{W}=0,+1$ for $A=1$.}
\end{figure}

\newpage
\begin{figure}
\epsfig{file=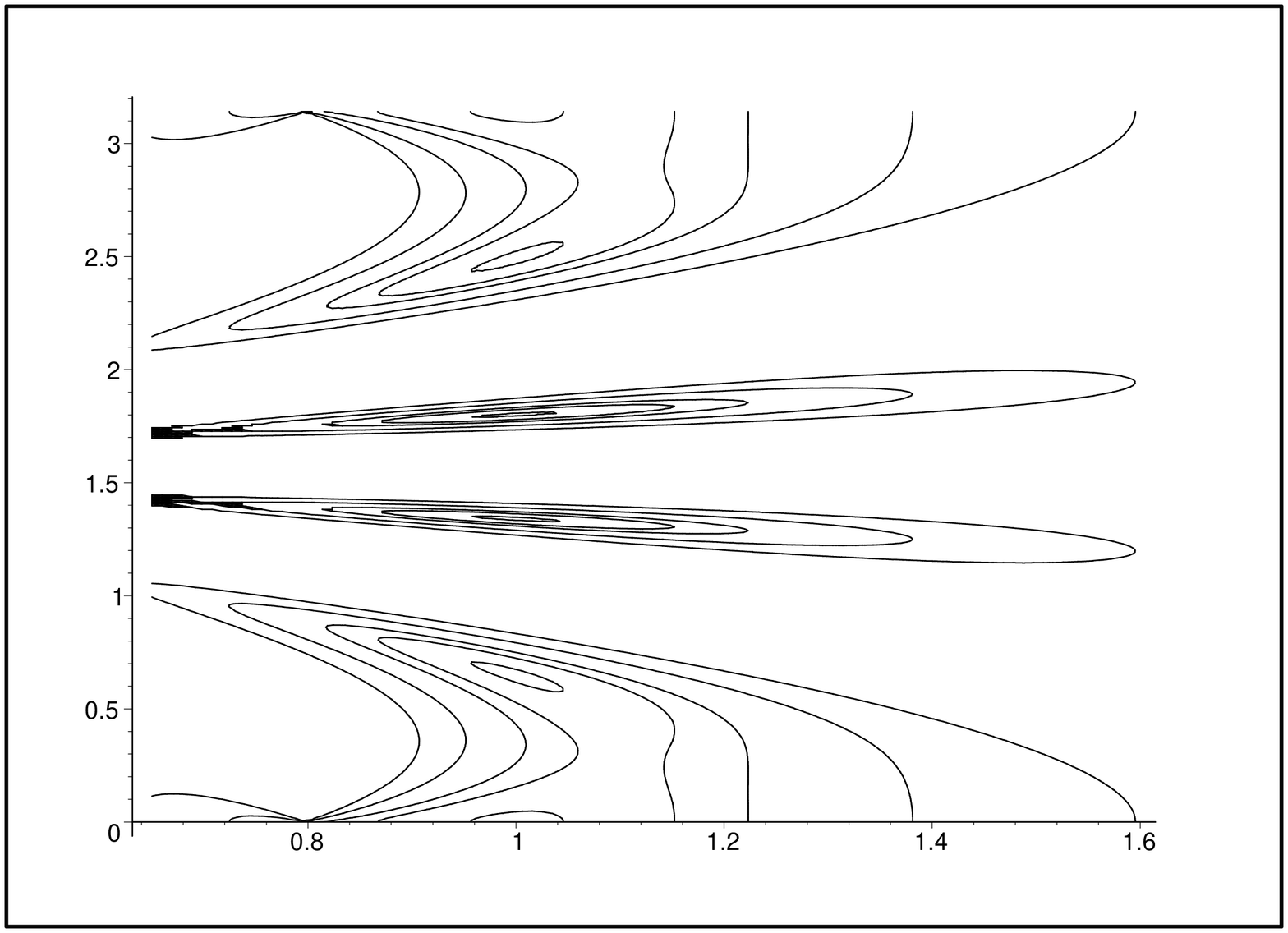,height=6in,angle=-90} \caption{Plot of
$ \mathcal{W}=0$ for $A=0.9, 0.95, 0.98, 0.99, 0.999$. Abscissa is
${\it x}$ and ordinate is $\theta$.  The enclosed area decreases
for increasing $A$ and vanishes for $A \geq 1$.}
\end{figure}

\end{document}